\documentclass[%
 aip, jcp,
 amsmath,amssymb,
 reprint,%
]{revtex4-2}
\usepackage{graphicx}
\usepackage{dcolumn}
\usepackage{bm}

\usepackage[utf8]{inputenc}
\usepackage[T1]{fontenc}
\usepackage{mathptmx}
\usepackage{etoolbox}
\usepackage{multirow}
\usepackage{braket}
\usepackage{hyperref}
\usepackage{cleveref}
\usepackage{xcolor}
\Crefname{equation}{Eq.}{Eqs.}
\Crefname{figure}{Fig.}{Figs.}
\Crefname{tabular}{Tab.}{Tabs.}
\makeatletter
\def\@email#1#2{%
 \endgroup
 \patchcmd{\titleblock@produce}
  {\frontmatter@RRAPformat}
  {\frontmatter@RRAPformat{\produce@RRAP{*#1\href{mailto:#2}{#2}}}\frontmatter@RRAPformat}
  {}{}
}%
\makeatother
\begin{document}

\title{Deterministic Optimisation of Jastrow Factors}
\author{Maria-Andreea Filip}
\email{maf63@cam.ac.uk}
\affiliation{Max Planck Institute for Solid State Research, Heisenbergstr. 1, 70569 Stuttgart, Germany}
\affiliation{Yusuf Hamied Department of Chemistry, University of Cambridge, Lensfield Road, Cambridge CB2 1EW, United Kingdom}
\author{Evelin Martine Corvid Christlmaier}
\author{J. Philip Haupt}
\author{Daniel Kats}
\author{Pablo López Ríos}
\affiliation{Max Planck Institute for Solid State Research, Heisenbergstr. 1, 70569 Stuttgart, Germany}
\author{Ali Alavi}
\email{a.alavi@fkf.mpg.de}
\affiliation{Max Planck Institute for Solid State Research, Heisenbergstr. 1, 70569 Stuttgart, Germany}
\affiliation{Yusuf Hamied Department of Chemistry, University of Cambridge, Lensfield Road, Cambridge CB2 1EW, United Kingdom}
\date{\today}

\begin{abstract}
    Highly flexible Jastrow factors have found significant use in stochastic electronic structure methods such as variational Monte Carlo (VMC) and diffusion Monte Carlo, as well as in quantum chemical  transcorrelated (TC) approaches, which have recently seen great success in generating highly accurate electronic energies using moderately sized basis sets. In particular for the latter, the intrinsic noise in the Jastrow factor due to its optimisation by VMC can pose a problem, especially when targeting weak (non-covalent) interactions. In this paper, we propose a deterministic alternative to VMC Jastrow optimisation, based on minimising the ``variance of the  TC reference energy" in a standard basis set. Analytic expressions for the derivatives of the TC Hamiltonian matrix elements are derived and implemented. This approach can be used to optimise the parameters in the Jastrow functions, either from scratch or to refine an initial VMC-based guess, to produce noise-free Jastrows in a reproducible manner. Applied to the first row atoms and molecules, the results show that the method yields Slater-Jastrow wavefunctions whose  variances are almost as low as those obtained from standard VMC variance optimisation, but whose energies are lower, and comparable to those obtained from energy-minimisation VMC. We propose that the method can be used both in the context of the transcorrelated method or in standard VMC as a new way to optimise Jastrow functions.
\end{abstract}
\maketitle

\section{Introduction}

Jastrow factors\cite{Jastrow1955} are commonly used in various \textit{ab-initio} electronic structure methods, where the introduction of explicit particle--particle correlation has been found to lead to much improved accuracy and convergence properties compared to standard approaches. In variational formulations, the use of such wavefunctions leads to many-body integrals,  the optimisation of which can only be done using stochastic Monte Carlo techniques. This has led to the development of the family of variational Monte Carlo methods \cite{Bartlett1955, Ceperley1977,Umrigar1988,Schmidt1990,Nightingale2001,Toulouse2007,Umrigar2007,Needs2020}, the output of which can be used as a trial wavefunction for the fixed-node diffusion Monte Carlo (DMC) procedure \cite{Anderson1975,Anderson1976,Ceperley1980,Foulkes2001}. 


There exists another class of explicitly correlated methods - the R12/F12 family of methods \cite{Hylleraas1928, Hylleraas1929, Hylleraas1930,Kutzelnigg1985,Klopper1987,Klopper1990,Kutzelnigg1991,Noga1994,Noga1997,Ten-no2004,Valeev2004,Kedzuch2005,Adler2007,Knizia2008,Knizia2009,Hattig2010,Kong2012}- in which the quantum chemical wavefunction is {\em additively} augmented with short-ranged fixed-amplitude geminal-type functions, designed mainly to capture electron-electron cusp conditions. In these approaches, many-body integrals (beyond 4-electron intergrals) are avoided, but the geminal functions themselves are typically not further optimised. These types of methods yield faster convergence to the basis-set limit of the underlying method (such MP2, CCSD),  but are difficult to extend beyond the double-excitation level of coupled-cluster theory \cite{Klopper1987, Klopper1990,Kutzelnigg1991,Valeev2004,Ten-no2004,Kedzuch2005,Knizia2008,Noga1994,Noga1997,Adler2007,Knizia2009,Hattig2010}. 


Transcorrelation is another distinct class of explicitly-correlated methods that share aspects of the previous two, and which offer in an interesting alternative. In this methodology, Jastrow factors are used  to similarity transform the Hamiltonian itself.\cite{Boys1969,Handy1969,Handy1971,Nooijen1998,Tenno2000,Tenno2000b,Hino2002,Sakuma2006,Tsuneyuki2008,Luo2018,Cohen2019} This leads to a non-hermitian Hamiltonian with 3-electron terms, approximate eigenvalues of which can be found by a variety of quantum chemical approaches,\cite{Cohen2019,Liao2021,Schraivogel2021,Ammar2022,Baiardi2022,Ammar2023,Ammar2023b,Liao2023,Schraivogel2023,Ammar2024, Morchen2025} provided they do not require a variational estimate of the energy. In general, this approach has been found to accelerate convergence to the basis-set limit and increase the single-reference character of the right eigenfunctions.\cite{Dobrautz2019, Haupt2023, Ammar2024} The use of very flexible Jastrow factors is a great advantage in this respect, albeit they necessitate numerical quadrature of the required matrix elements. Thanks to normal-ordering approximations \cite{Liao2021, Schraivogel2021} such as the xTC approximation \cite{Christlmaier2023}, the TC hamiltonian can be reduced to two-electron interactions, greatly decreasing the burden of additional TC integral calculations.   

Both the DMC approach and many modern transcorrelated (TC) methods employ parametrised Jastrow factors,\cite{Boys1969, Drummond2004} which must be optimised. In both cases, this is usually done using VMC to minimise either the variational energy or the variance of the energy of the Slater--Jastrow wavefunction.\cite{Umrigar1988,Kent1999} Stochastic optimisation has proved quite effective to treat the often large number of parameters involved in this type of Jastrow factor, but induces unavoidable noise in the final values. This is not an issue for the DMC energy, which is independent of the Jastrow factor, but may pose problems for TC methods, especially when small energy difference are being sought which may be overwhelmed by the stochastic noise arising from the Jastrow factor.

An interesting question for the TC methodology is that, since it avoids many-electron integrals (beyond 3 electron terms)  in the first place,  can it be used to {\em deterministically} optimise the Jastrow parameters? Of course, due attention needs to be paid to the fact that, as a non-hermitian problem, the TC energy itself cannot be used as a target function to optimise, and a target function based on a variance quantity must be used instead. In this work, we present such an approach, based on the minimisation of the ``variance of the TC reference energy" often employed in TC approaches.\cite{Haupt2023} We show that, when applied after VMC, this approach lowers the variance of energies obtained from different Jastrow optimisation runs, with resulting transcorrelated energies maintaining similar accuracy to those obtained from pure VMC-optimised Jastrow factors. Furthermore, VMC energies themselves are shown to be lower than those obtained by stochastic minimisation of the variance of the reference energy, while maintaining lower variances than those obtained by conventional VMC energy minimisation.

\section{Transcorrelated Methodology}

\subsection{Similarity Transform of the Hamiltonian}

Rather than using the Slater-Jastrow form of the wavefunction,
\begin{equation}
    \ket{\Psi} = e^\tau \ket{\Phi},
\end{equation}
as an Ansatz to be directly optimised, in transcorrelated methods one uses the Jastrow factor to similarity transform the Hamiltonian:
\begin{equation}
   \hat H_\text{TC} = e^{-\tau} \hat H e^\tau.
\end{equation}
which can then be used to calculate $|\Phi\rangle$ as the right-eigenfunction of $\hat H_\text{TC}$. If the Jastrow  function depends purely on the spatial coordinates of the particles, this Hamiltonian can be expressed as a Baker-Campbell-Hausdorff expansion, which terminates at the second nested commutator:
\begin{equation}
   \hat H_\text{TC} = \hat H + [\hat H, \tau] + \frac{1}{2} [[\hat H, \tau]\tau]. 
\end{equation}

Consider a Jastrow factor that contains no more than two-electron interactions, 
\begin{equation}
    \tau = \sum_{i < j} u(\mathbf{r}_i, \mathbf{r}_j),
    \label{eq:jas}
\end{equation}
where $u$ is a symmetric correlation function. The TC Hamiltonian may then be rewritten as
\begin{equation}
  \hat H_\mathrm{TC}[\tau] = \hat H - \hat K[\tau] - \hat L[\tau],   
\end{equation}
where we have emphasised the dependence of the additional terms on the Jastrow factor,
\begin{align}
\begin{split}
   \hat K(\mathbf{r}_i, \mathbf{r}_j)= \frac{1}{2}&[\nabla_i^2u(\mathbf{r}_i, \mathbf{r}_j)+\nabla_j^2u(\mathbf{r}_i, \mathbf{r}_j)\\
    +&|\nabla_iu(\mathbf{r}_i, \mathbf{r}_j)|^2+|\nabla_ju(\mathbf{r}_i, \mathbf{r}_j)|^2]\\
    +&\nabla_iu(\mathbf{r}_i, \mathbf{r}_j)\cdot \nabla_i+\nabla_ju(\mathbf{r}_i, \mathbf{r}_j)\cdot \nabla_j
\end{split}
\label{eq:K}
\end{align}
and
\begin{align}
    \begin{split}
        \hat L(\mathbf{r}_i, \mathbf{r}_j, \mathbf{r}_k) = &\nabla_iu(\mathbf{r}_i, \mathbf{r}_j)\cdot \nabla_iu(\mathbf{r}_i, \mathbf{r}_k)\\
        + &\nabla_ju(\mathbf{r}_j, \mathbf{r}_k)\cdot \nabla_ju(\mathbf{r}_j, \mathbf{r}_i)\\
        + &\nabla_ku(\mathbf{r}_k, \mathbf{r}_i)\cdot \nabla_ku(\mathbf{r}_k, \mathbf{r}_j)
    \end{split}
    \label{eq:L}
\end{align}

\subsection{Stochastic Jastrow Optimisation}

In most modern TC methods, $\hat H_\mathrm{TC}$ is obtained by optimising a parametrised $u$ with respect to some reference wavefunction $\ket{\Phi}$, which may be either the conventional Hartree--Fock determinant or a short configuration interaction (CI) expansion.\cite{haupt2025} This is done using VMC and minimising one of
\begin{itemize}
    \item the variational energy of the Slater--Jastrow wavefunction
    \begin{equation}
        \braket{E} = \frac{\braket{\Phi|(e^\tau)^\dagger \hat H e^\tau|\Phi}}{\braket{\Phi|(e^\tau)^\dagger e^\tau|\Phi}}
    \end{equation}
    \item the variance of the energy of the Slater--Jastrow wavefunction    
    \begin{equation}
        \sigma^2 = \frac{\braket{\Phi|(e^\tau)^\dagger (\hat H - E)^2 e^\tau|\Phi}}{\braket{\Phi|(e^\tau)^\dagger e^\tau|\Phi}}
    \end{equation}
    \item the ``variance of the reference energy"\cite{Haupt2023}\footnote{The expression given here is a corrected version of Eq. 19 in Ref. \onlinecite{Haupt2023}, to agree with Eq. 20 of the same publication, which is the relevant VMC working equation.}
    \begin{equation}
        \sigma_\mathrm{ref}^2 = \braket{\Phi||\hat H_\mathrm{TC} - E_\mathrm{ref}|^2 |\Phi},
        \label{eq:var_ref}
    \end{equation}
    where $E_\mathrm{ref} = \braket{\Phi|\hat H_\mathrm{TC}  |\Phi}$ is the (TC) reference energy.
\end{itemize}
The latter of these quantities was recently shown to provide the best compromise between value and spread of the VMC energy and has therefore been proposed as the standard approach for TC calculations.\cite{Haupt2023} In VMC, it can be computed as the sample variance of the local Slater--Jastrow energy over the reference distribution
\begin{equation}
    S^2_\mathrm{ref} = \frac{1}{n_\mathrm{opt}-1}\sum_{n=1}^{n_\mathrm{opt}}\Big|\frac{\hat H \Psi(\mathbf{R}_n)}{\Psi(\mathbf{R}_n)} - \widetilde E_\mathrm{ref}\Big|^2
    \label{eq:variance_reference_vmc}
\end{equation}
where $\{\mathbf{R}_n\}$ are electron distributions sampled from $|\Phi|^2$ and 
\begin{equation}
    \widetilde E_\mathrm{ref}= \frac{1}{n_\mathrm{opt}}\sum_{n=1}^{n_\mathrm{opt}}\frac{\hat H \Psi(\mathbf{R}_n)}{\Psi(\mathbf{R}_n)}. 
\end{equation}

Two types of parametrised Jastrow factors have been used recently in TC applications:
\begin{itemize}
    \item Boys--Handy (BH) Jastrow factors\cite{Boys1969} of the form
    \begin{equation}
        u(\mathbf{r}_i, \mathbf{r}_j, \mathbf{r}_I) = \sum_{m,n,o}^{m+n+o \leq 6} c_{mno}\bar r_{iI}^m \bar r_{jI}^n \bar r_{ij}^o,
    \end{equation}
    where $\bar r$ is a scaled distance. This can take various forms, the most common of which is $\bar r = \frac{r}{1+r}$.
    \item Drummond--Towler--Needs (DTN) Jastrow factors\cite{Drummond2004,Lopez2012} of the form
    \begin{equation}
    \begin{split}
        J(\mathbf{r}_i, \mathbf{r}_j,\mathbf{r}_I)  &= \sum_{i<j}u(r_{ij}) + \sum_{i,I}\chi(r_{iI})\\
        &+ \sum_{i<j,I}\zeta(r_{ij}, r_{iI}, r_{jI}),
    \end{split}
\end{equation}
with the electron-electron, electron-nucleus and electron-electron nucleus terms expressed as natural power expansions
\begin{equation}
    u(r_{ij}) = t(r_{ij},L_u)\sum_k^{N_u} a_k r_{ij}^k,
    \label{eq:u}
\end{equation}
\begin{equation}
    \chi(r_{iI}) = t(r_{iI},L_\chi)\sum_k^{N_\chi} b_k r_{iI}^k
    \label{eq:chi}
\end{equation}
and
\begin{equation}
   \begin{split}
       \zeta(r_{ij}, r_{iI}, r_{jI}) &= t(r_{iI},L_\zeta)t(r_{jI},L_\zeta)\times\\
   &\times \sum_{klm}^{N_f} c_{klm} r_{ij}^kr_{iI}^lr_{jI}^m,
   \end{split}
    \label{eq:f}
\end{equation}

where $\{a_k\}$, $\{b_k\}$, $\{c_{klm}\}$ are linear coefficients, ${t(r,L)=(1-r/L)^3\Theta(L-r)}$, $\Theta(x)$ is the Heaviside function and $L_u$, $L_\chi$, $L_\zeta$ are some appropriate cutoff radii.
\end{itemize}
As the the nucleus positions are kept fixed, the terms of the DTN Jastrow can be combined as
\begin{align}
\begin{split}
    \frac{1}{2}\sum_{i\neq j}^N (u_{ij}+\zeta_{ij}) + \sum_i^N \chi_i &= \frac{1}{2}\sum_{i \neq j}^N (u_{ij} + \zeta_{ij}) + \frac{1}{2}(\sum_i^N \chi_i +\sum_j^N \chi_j)\\
    &=\frac{1}{2}\sum_{i\neq j}^N\left(u_{ij} +\zeta_{ij} + \frac{1}{(N-1)}\left(\chi_i + \chi_j\right)\right),
\end{split}
\end{align}
where we have used the notation $u_{ij} = u(r_{ij})$, $\chi_i = \sum_I \chi(r_{iI})$ and $\zeta_{ij} = \sum_I \zeta(r_{ij}, r_{iI}, r_{jI})$. Therefore both Jastrow forms above can be rewritten in the general form of \Cref{eq:jas}, with
\begin{equation}
    u(\mathbf{r}_i, \mathbf{r}_j) = \sum_l f_l u_l (\mathbf{r}_i, \mathbf{r}_j).
    \label{eq:form}
\end{equation}
For the BH Jastrow, $l$ is a counter over the different $(m,n,o)$ contributions, $f_l = c_{mno}$ and $u_l(\mathbf{r}_i, \mathbf{r}_j) = \bar r_{iI}^m \bar r_{jI}^n \bar r_{ij}^o$ for the corresponding $(m,n,o)$. For the DTN Jastrow, $f_l$ is one of $a_k,\ b_k,\ c_{klm}$ and $u_l$ is the corresponding function, $t(r_{ij}, L_u)r_{ij}^k$, ${\sum_I \left(t(r_{iI}, L_\chi) r_{iI}^k + t(r_{jI}, L_\chi) r_{jI}^k\right)}/{(N-1)}$ or $t(r_{iI},L_\zeta)t(r_{jI},L_\zeta) c_{klm} r_{ij}^kr_{iI}^lr_{jI}^m$. For convenience, we will assume the general form in \Cref{eq:form} when deriving the gradients for the deterministic optimisation of the Jastrow factors, although a similar derivation can be carried out without folding the one-electron terms of the Jastrow into the two-electron term.

\subsection{Integral Evaluation}

Given a particular Jastrow, the resulting TC Hamiltonian may be written in spin-orbital notation as
\begin{equation}
\begin{split}
    \hat H_\text{TC} &= \sum_{PQ} h_P^{Q}\hat a_P^\dagger \hat a_Q + \frac{1}{2}\sum_{PQRS} (V_{PR}^{QS}-K_{PR}^{QS})\hat a_P^\dagger \hat a_R^\dagger  \hat a_S \hat a_Q\\
    &- \frac{1}{6} \sum_{PQRSTU} L^{QSU}_{PRT}\hat a_P^\dagger \hat a_R^\dagger \hat a_T^\dagger  \hat a_U \hat a_S \hat a_Q ,
    \end{split}
\end{equation}
for a set of one-particle basis functions $\{\phi_P\}$, where $h_{P}^{Q} = -\frac{1}{2}\braket{\phi_P|\nabla^2|\phi_Q} - \sum_I Z_I \braket{\phi_P|\mathbf{r}_I^{-1}|\phi_Q}$, $V_{PR}^{QS} =\braket{\phi_P\phi_R|\mathbf{r}_{12}^{-1}|\phi_Q\phi_S}$, ${K_{PR}^{QS} = \braket{\phi_P\phi_R|\hat K|\phi_Q\phi_S}}$ and ${L_{PRT}^{QSU} = \braket{\phi_P\phi_R\phi_T|\hat L|\phi_Q\phi_S\phi_U}}$. One can compute the one-, two- and three-body integrals of the Hamiltonian and use them in subsequent post-Hartree--Fock treatments.
The three-body operator $\hat L$ is the bottleneck of this process, but it has been shown that excluding the explicit three-body contributions ($L_{PRT}^{QSU}$ where all 6 indices are distinct) has little effect on the resulting energies.\cite{Schraivogel2021,Schraivogel2023} The remaining three-body terms can be folded into the lower-order integrals, significantly reducing the cost of computing the Hamiltonian, in what is known as the xTC approximation.\cite{Christlmaier2023} This leads to a correction to the two-body term given by
\begin{equation}
    \Delta U_{PR}^{QS} = - \sum_{TU} (L_{PRT}^{QSU} - L_{PRT}^{QUS} - L_{PRT}^{USQ})\gamma_U^T,
    \label{eq:delU}
\end{equation}
where $\gamma_U^T =\braket{\Phi|\hat a_T^\dagger \hat a_U|\Phi}$ are elements of the density matrix of the reference wavefunction $\Phi$. For a single-determinant reference, the one-body correction can then be computed as
\begin{equation}
    \Delta h_P^Q = -\frac{1}{2}\sum_{RS} (\Delta U_{PR}^{QS} - \Delta U_{PR}^{SQ})\gamma_S^R,
\end{equation}
and the zero-body correction needed to account for the normal ordering of the Hamiltonian as
\begin{equation}
    \braket{\Phi|L|\Phi} = - \frac{1}{3} \sum_{PQ} \Delta h_P^Q \gamma_Q^P.
\end{equation}

\section{Deterministic Jastrow Optimisation}
Access to the TC Hamiltonian integrals suggests a new path to Jastrow optimisation that does not require a stochastic approach. If we consider the variance of the reference energy in \Cref{eq:var_ref}, it may be rewritten as\cite{Haupt2023}
\begin{align}
        \sigma_\mathrm{ref}^2 &= \braket{\Phi|(\hat H_\mathrm{TC} - E_\mathrm{ref})^\dagger (\hat H_\mathrm{TC} - E_\mathrm{ref})|\Phi}\\
        &= \sum_I \braket{\Phi|(\hat H_\mathrm{TC} - E_\mathrm{ref})^\dagger|\Phi_I}\braket{\Phi_I|(\hat H_\mathrm{TC} - E_\mathrm{ref})|\Phi},
    \label{eq:var_resolution}
\end{align}
where $\ket{\Phi_I}$ are Slater determinants spanning the complete problem Hilbert space. In the particular case where the reference wavefunction is the Hartree--Fock determinant $\ket{\Phi_0}$, \Cref{eq:var_resolution} becomes
\begin{equation}
    \sigma_\mathrm{ref}^2 = \sum_{I\neq 0} \braket{\Phi_I|\hat H_\mathrm{TC}|\Phi_0}^2
    \label{eq:var_resolution_0}
\end{equation}
When $\ket{\Phi_I}$ only span a finite basis set, \Cref{eq:var_resolution_0} provides an approximation for the variance of the reference energy which can be easily computed using the TC Hamiltonian integrals. We therefore propose using this approximate quantity as a cost function for Jastrow optimisation and in the following sections investigate the requirements on the basis set to make this method viable, as well as its applicability in VMC and  TC calculations. 
\subsection{Analytical Gradients}
An advantage of \Cref{eq:var_resolution_0} as an optimisation target is that analytical gradients can be easily computed using the same paradigm as for the matrix elements themselves. The gradients of the variance with respect to Jastrow parameters are given by
\begin{equation}
    \frac{d\sigma_\mathrm{ref}^2}{df_l} = \sum_{I\neq 0} 2\braket{\Phi_I|\hat H_\mathrm{TC}|\Phi_0}\frac{d}{df_l}\braket{\Phi_I|\hat H_\mathrm{TC}|\Phi_0}.
    \label{eq:var_resolution_0_grad}
\end{equation}
If we are interested in obtaining matrix elements of the form
\begin{equation}
    \frac{\partial}{\partial f_l} \langle \Phi_I|H_\mathrm{TC}|\Phi_0\rangle = -\langle \Phi_I| \frac{\partial (K+L) }{\partial f_l}|\Phi_0\rangle,
\end{equation}
we are required to obtain the corresponding derivative integrals of the form $\braket{PR|\frac{\partial K}{\partial f_l}|QS}$ and $\braket{PRT|\frac{\partial L}{\partial f_l}|QSU}$.
\begin{table*}[]
    \centering
    \bgroup
\def\arraystretch{2}
    \begin{tabular}{l|l||l|l}
        \multicolumn{1}{c|}{$\Delta U_{PR}^{QS}$} &  \multicolumn{1}{c||}{$d\Delta U_{PR}^{QS}/df_l$}&\multicolumn{1}{c|}{$\Delta U_{PR}^{QS}$}&\multicolumn{1}{c}{$d\Delta U_{PR}^{QS}/df_l$}\\
        \hline
         $\rho_P^Q(i) =w_i \phi^*_P(i)\phi_Q(i)   $&  &$A_R^S(i) = \tilde V_R^S(i) - \tilde Z _R ^S (i)   $ & $   \frac{d A_R^S(i)}{df_l} = \frac{d \tilde{V}_R^S(i)}{df_l} - \frac{d \tilde{Z}_R^S(i)}{df_l}$\\
        $\mathbf{W}_i = \mathbf{V}_T^U(i)\gamma_U^T   $ & $   \frac{d\mathbf{W}_i}{df_l} = \frac{d\mathbf{V}_T^U(i)}{df_l}\gamma_U^T$& $\mathbf{Y}_R^T(i) = \mathbf{V}_R^U(i)\gamma_U^T $ & $\frac{d \mathbf{Y}_R^T(i)}{df_l} = \frac{d\mathbf{V}_R^U(i)}{df_l}\gamma_U^T$\\
        $\tilde{V}_R^S(i) = \mathbf{W}_i \cdot \mathbf{V}_R^S(i)   $ & $   \frac{d \tilde{V}_R^S(i)}{df_l} = \frac{d\mathbf{W}_i}{df_l}\cdot \mathbf{V}_R^S(i) + \mathbf{W}_i \cdot \frac{d\mathbf{V}_R^S(i)}{df_l}$&$\mathbf{Z}_R^S(i) = \rho_R^U\mathbf{X}_U^S(i) + \mathbf{Y}_R^T(i)\rho_T^S(i)$ & $\frac{d \mathbf{Z}_R^S(i)}{df_l} = \rho_R^U\frac{d \mathbf{X}_U^S(i)}{df_l}+ \frac{d \mathbf{Y}_R^T(i)}{df_l}\rho_T^S(i)$\\
        $\mathbf{X}_U^S(i) = \mathbf{V}_T^S(i)\gamma_U^T   $ & $   \frac{d \mathbf{X}_U^S(i)}{df_l} = \frac{d\mathbf{V}_T^S(i)}{df_l}\gamma_U^T$&$\tilde W(i) = \rho_T^U(i)\gamma_U^T$ & \\
        $\tilde{Z}_R^S(i) =  \mathbf{V}_R^U(i) \cdot  \mathbf{X}_U^S(i)   $ & $   \frac{d \tilde{Z}_R^S(i)}{df_l} = \mathbf{V}_R^S(i) \cdot \frac{d \mathbf{X}_U^S(i)}{df_l}+  \frac{d\mathbf{V}_R^S(i)}{df_l}\cdot \mathbf{X}_U^S(i)$&        $\mathbf{B}_R^S(i) = \frac{1}{2} \tilde W(i) \mathbf{V}_R^S(i) - \mathbf{Z}_R^S(i)$ & $\frac{d \mathbf{B}_R^S(i)}{df_l} =  \frac{1}{2} \tilde W(i) \frac{\mathbf{V}_R^S(i)}{df_l} - \frac{d \mathbf{Z}_R^S(i)}{df_l}$\\
        \hline
    \end{tabular}
    \egroup
    \caption{Derivatives with respect to $f_l$ of various xTC intermediates, as given in Ref.\onlinecite{Christlmaier2023}.}
    \label{tab:xtc_deriv}
\end{table*}

Given the forms of $\hat K$ and $\hat L$ given in \Cref{eq:K,eq:L}, there are four types of terms which must be differentiated with respect to $f_l$:
\begin{align}
K_1(\mathbf{r}_1, \mathbf{r}_2) &= \nabla_1^2u\\
K_2(\mathbf{r}_1, \mathbf{r}_2) &= (\nabla_1u)^2\\
K_3(\mathbf{r}_1, \mathbf{r}_2) &= \nabla_1u\cdot\nabla_1\\
L_1(\mathbf{r}_1, \mathbf{r}_2, \mathbf{r}_3) &= \nabla_1 u(\mathbf{r}_1, \mathbf{r}_2))\cdot \nabla_1 u(\mathbf{r}_1, \mathbf{r}_3).
\end{align}
In practice, the integrals of $K_1$ are computed by parts, so it takes a form similar to $K_3$. However
\begin{align}
    \begin{split}
\langle PR | K_3 | QS \rangle &= \int\int \phi_P(\mathbf{r}_1) \phi_R(\mathbf{r}_2) \nabla_1u\cdot\nabla_1[\phi_Q(\mathbf{r}_1) \phi_S(\mathbf{r}_2)] d\mathbf{r}_1 d\mathbf{r}_2 \\
&=  \int\int \phi_P(\mathbf{r}_1) \phi_R(\mathbf{r}_2) \phi_Q(\mathbf{r}_2) \nabla_1u\cdot\nabla_1\phi_S(\mathbf{r}_1)  d\mathbf{r}_1 d\mathbf{r}_2
    \end{split}
\end{align}
while 
\onecolumngrid

\noindent\rule[0.5ex]{\linewidth}{1pt}
\begin{align}
    \begin{split}
\langle PR | K_1 | QS \rangle &= \int\int \phi_P(\mathbf{r}_1) \phi_R(\mathbf{r}_2) (\nabla_1^2u)\phi_Q(\mathbf{r}_1) \phi_S(\mathbf{r}_2) d\mathbf{r}_1 d\mathbf{r}_2=  - \int\int  \nabla_1u\cdot\nabla_1\left[\phi_P(\mathbf{r}_1) \phi_R(\mathbf{r}_2) \phi_Q(\mathbf{r}_1)\phi_S(\mathbf{r}_2)\right]  d\mathbf{r}_1 d\mathbf{r}_2 \\
& = - \int\int \phi_R(\mathbf{r}_2)\phi_S(\mathbf{r}_2) \nabla_1u\cdot\left[\phi_P(\mathbf{r}_1) \nabla_1\phi_Q(\mathbf{r}_1) + \phi_Q(\mathbf{r}_1) \nabla_1\phi_P(\mathbf{r}_1)\right] d\mathbf{r}_1 d\mathbf{r}_2
    \end{split}
\end{align}
\noindent\rule[0.5ex]{\linewidth}{1pt}
\twocolumngrid
The relevant derivatives of the Jastrow factor are therefore
\begin{align}
    \begin{split}
 \frac{\partial (\nabla_1u)^2 }{\partial f_l} &= 2 \nabla_1 u \cdot \frac{\partial \nabla_1u }{\partial f_l} \\&= 2 \nabla_1 u \cdot\frac{\partial} {\partial f_l} \nabla_1\sum_j f_j u_j = \nabla_1 u \cdot\nabla_1 u_l\\
    \end{split}
\end{align}
\begin{align}
    \begin{split}
    \frac{\partial}{\partial f_l} (\nabla_1u)\cdot\nabla_1 &=\frac{\partial}{\partial f_l} (\nabla_1\sum_j f_j u_j)\cdot\nabla_1 \\
    &= \frac{\partial}{\partial f_l} \sum_j f_j \nabla_1 u_j\cdot\nabla_1 = \nabla_1 u_l \cdot\nabla_1\\
    \end{split}
    \end{align}
\begin{align}
\begin{split}
\frac{\partial L_1}{\partial f_l} &=\frac{\partial}{\partial f_l} (\nabla_1\sum_j f_j u_j (\mathbf{r}_1, \mathbf{r}_2))\cdot (\nabla_1\sum_k f_k u_k(\mathbf{r}_1, \mathbf{r}_3)) \\
&= \frac{\partial}{\partial f_l} (\sum_j f_j \nabla_1 u_j (\mathbf{r}_1, \mathbf{r}_2))\cdot (\sum_k f_k \nabla_1u_k(\mathbf{r}_1, \mathbf{r}_3)) \\
&= \nabla_1 u_l (\mathbf{r}_1, \mathbf{r}_2))\cdot \nabla_1 u (\mathbf{r}_1, \mathbf{r}_3)) + \nabla_1 u (\mathbf{r}_1, \mathbf{r}_2))\cdot \nabla_1 u_l (\mathbf{r}_1, \mathbf{r}_3))
    \end{split}
\end{align}

Compared to an energy evaluation, the only new terms needed are $\nabla_{1} u_l$ and $\nabla_{2} u_l$. In practice, these terms are summed to obtain $\nabla_{1} u$ and $\nabla_{2} u$, so they are easily accessible.

In the TCHInt package\cite{tchint} used for TC calculations, the $K$ integrals are computed in two steps, the first being to compute the intermediate
\begin{align}
\begin{split}
    x_{P}^{Q}(\mathbf{r}_2)= - & \int\nabla_1u\cdot[\phi_P(\mathbf{r}_1) \nabla_1\phi_Q(\mathbf{r}_1) + \phi_Q(\mathbf{r}_1) \nabla_1\phi_P(\mathbf{r}_1)] d\mathbf{r}_1\\  + & \int\phi_P(\mathbf{r}_1)\nabla u^2 \phi_Q(\mathbf{r}_1)d\mathbf{r}_1 +2\int\phi_P(\mathbf{r}_1)\nabla u\cdot\nabla \phi_Q(\mathbf{r}_1) d\mathbf{r}1 \\
    = &\int\nabla_1u\cdot[\phi_{P}(\mathbf{r}_1) \nabla_1\phi_Q(\mathbf{r}_1) - \phi_Q(\mathbf{r}_1) \nabla_1\phi_P(\mathbf{r}_1)] d\mathbf{r}_1\\
    +  &\int\phi_P(\mathbf{r}_1)\nabla u^2 \phi_Q(\mathbf{r}_1)d\mathbf{r}_1,
\end{split}
\end{align}
followed by
\begin{equation}
\begin{split}
    K_{PR}^{QS} &= \frac{1}{2}\Big[\int\phi_R(\mathbf{r}_2)x_{P}^{Q}(\mathbf{r}_2)\phi_S(\mathbf{r}_2)d\mathbf{r}_2 \\
&+ \int\phi_P(\mathbf{r}_2)x_{R}^{S}(\mathbf{r}_2)\phi_Q(\mathbf{r}_2)d\mathbf{r}_2\Big].
\end{split}
\end{equation}
$L$ is computed by first obtaining
\begin{equation}
\mathbf{V}_{P}^{Q}(\mathbf{r}) = \int\nabla u(\mathbf{r},\mathbf{r}_2) \phi_P(\mathbf{r}_2) \phi_Q(\mathbf{r}_2) d\mathbf{r}_2    
\end{equation}
and then
\begin{equation}
    \begin{split}
         L_{PRT}^{QSU} &= \int \phi_T(\mathbf{r})\phi_U(\mathbf{r})\mathbf{V}_{P}^{Q}(\mathbf{r})\cdot\mathbf{V}_{R}^{S}(\mathbf{r})  \\
         &+\int \phi_R(\mathbf{r})\phi_S(\mathbf{r})\mathbf{V}_{T}^{U}(\mathbf{r})\cdot\mathbf{V}_{P}^{Q}(\mathbf{r})\\
         &+\int \phi_P(\mathbf{r})\phi_Q(\mathbf{r})\mathbf{V}_{R}^{S}(\mathbf{r})\cdot\mathbf{V}_{T}^{U}(\mathbf{r}).
    \end{split}
\end{equation}

The same pattern may be followed to obtain the corresponding derivatives, computing first

\begin{align}
    \begin{split}
       \frac{dx_{P}^{Q}(\mathbf{r}_2)}{df_l} 
        = -\frac{1}{2}&\int \nabla_1u_l\cdot\nabla_1(\phi_P(\mathbf{r}_1)\phi_Q(\mathbf{r}_1)) \mathrm{d}\mathbf{r}_1\\+  &\int \phi_P(\mathbf{r}_1)\nabla_1u_l\cdot\nabla_1\phi_Q(\mathbf{r}_1) \mathrm{d}\mathbf{r}_1 \\+
        &\int \phi_P(\mathbf{r}_1)\nabla_1u_l\cdot\nabla_1u\phi_Q(\mathbf{r}_1)\mathrm{d}\mathbf{r}_1 \\
        = \frac{1}{2}\Big[&\int \phi_P(\mathbf{r}_1)\nabla_1u_l\cdot\nabla_1\phi_Q(\mathbf{r}_1) \mathrm{d}\mathbf{r}_1 \\
        - &\int \phi_Q(\mathbf{r}_1)\nabla_1u_l\cdot\nabla_1\phi_P(\mathbf{r}_1) \mathrm{d}\mathbf{r}_1 \Big] \\ +
        &\int \phi_P(\mathbf{r}_1)\nabla_1u_l.\nabla_1u\phi_Q(\mathbf{r}_1)\mathrm{d}\mathbf{r}_1 
    \end{split}
\end{align}
\begin{figure*}
    \includegraphics[width=\textwidth]{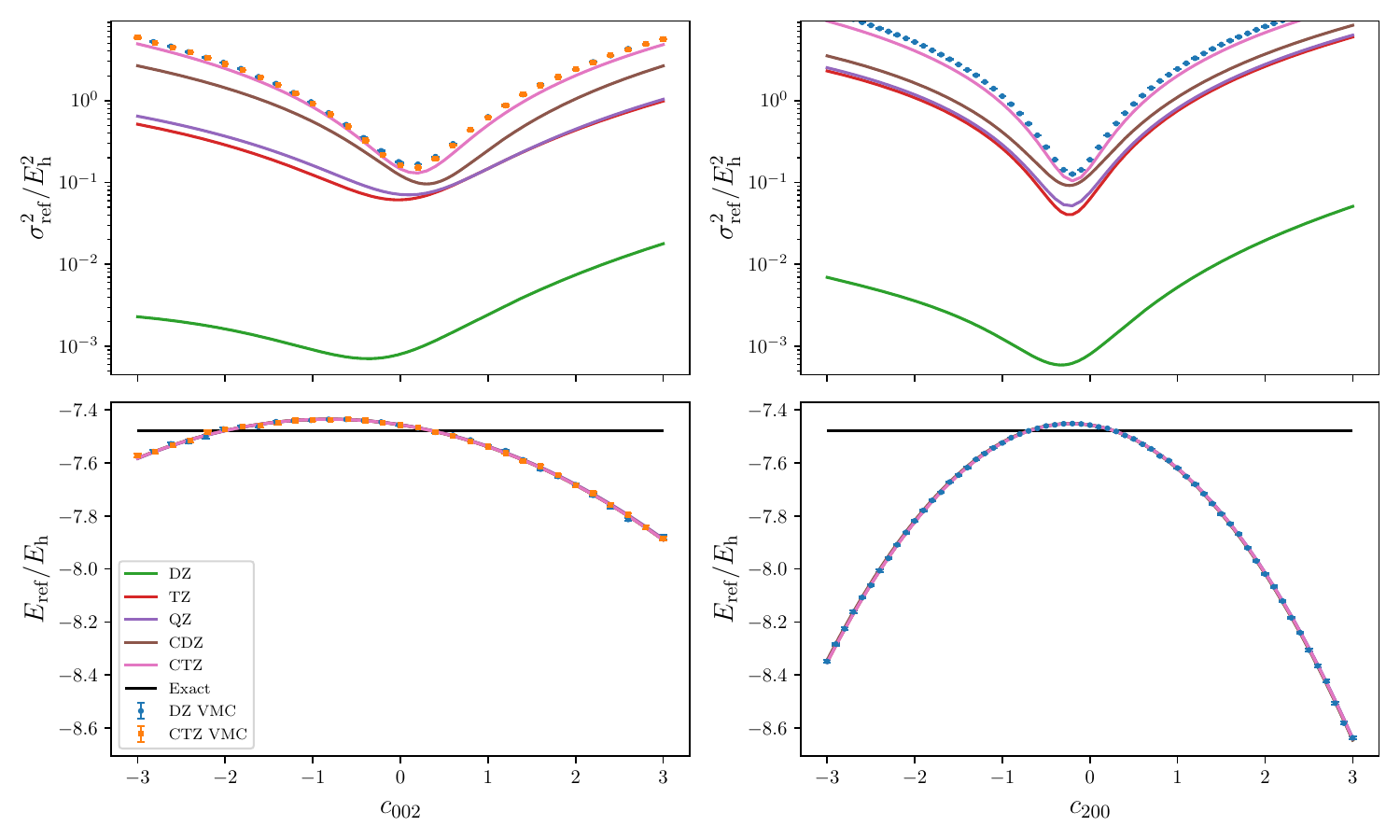}
    \caption{The variance of the reference energy (top) and the reference energy itself (bottom) for the Li atom, as a function of the parameter controlling the first free electron-electron (left, $m = n = 0$, $o = 2$) and electron-nucleus (right, $m(n) = 2$, $n(m) = o = 0$) correlation term in a BH Jastrow, with all other optimizable parameters fixed at zero. The quantities are computed using different basis sets, both deterministically (solid lines) and with VMC (dotted lines). The cc-pCVTZ VMC curves are only reported for the left plots. Deterministic reference energies for all basis sets overlap to within the resolution of the plots.}
    \label{fig:one_param}
\end{figure*}
and
\begin{equation}
  \frac{d\mathbf{V}_{P}^Q(\mathbf{r}_2)}{df_l}= \int\nabla_1 u_l(\mathbf{r}_1,\mathbf{r}_2) \phi_P(\mathbf{r}_1) \phi_Q(\mathbf{r}_1) d\mathbf{r}_1.  
\end{equation}

These can then be used to obtain
\begin{equation}
\begin{split}
    \frac{dK_{PR}^{QS}}{df_l} &= \int\phi_R(\mathbf{r}_2)\frac{dx_P^Q}{df_l}(\mathbf{r}_2)\phi_S(\mathbf{r}_2)d\mathbf{r}_2 \\
    &+ \int\phi_P(\mathbf{r}_2)\frac{dx_R^S}{df_l}(\mathbf{r}_2)\phi_Q(\mathbf{r}_2)d\mathbf{r}_2 
\end{split}
\end{equation}
and
\begin{align}
    \begin{split}
\frac{dL_{PRT}^{QSU}}{df_l} = &+\int \phi_T(\mathbf{r})\phi_U(\mathbf{r})\frac{d\mathbf{V}_{P}^{Q} (\mathbf{r})}{df_l}\cdot\mathbf{V}_{R}^{S}(\mathbf{r}) \\
&+\int \phi_T(\mathbf{r})\phi_U(\mathbf{r})\mathbf{V}_{P}^{Q}(\mathbf{r})\cdot \frac{d\mathbf{V}_{R}^{S} (\mathbf{r}) }{df_l} \\
&+\int \phi_R(\mathbf{r})\phi_S(\mathbf{r})\frac{d\mathbf{V}_{T}^{U}(\mathbf{r})}{df_l}
\cdot\mathbf{V}_{P}^{Q}(\mathbf{r}) \\
&+\int \phi_R(\mathbf{r})\phi_S(\mathbf{r})\mathbf{V}_{T}^{U}(\mathbf{r})\cdot \frac{d\mathbf{V}_{P}^{Q}(\mathbf{r})}{df_l}\\ 
&+\int \phi_P(\mathbf{r})\phi_Q(\mathbf{r})\frac{d\mathbf{V}_{R}^{S}(\mathbf{r})}{df_l}\cdot\mathbf{V}_{T}^{U}(\mathbf{r}) \\
&+\int \phi_P(\mathbf{r})\phi_Q(\mathbf{r})\mathbf{V}_{R}^{S}(\mathbf{r})\cdot \frac{d\mathbf{V}_{T}^{U}(\mathbf{r})}{df_l}
    \end{split}
\end{align}

\subsection{xTC Derivatives}
In practice, the two-body correction $\Delta U_{PR}^{QS}$ in \Cref{eq:delU} is computed from a series of intermediates shown in \Cref{tab:xtc_deriv}, as\cite{Christlmaier2023}
\begin{equation}
    \Delta U_{PR}^{QS} = - \mathcal{P}_{(RS)}^{(PQ)}\left(\rho^Q_P (i)A^S_R(i) + \mathbf{V}^Q_P(i) \cdot \mathbf{B}^S_R (i) \right),
\end{equation}
The intermediates needed to compute the derivative of the two-body correction $\Delta U_{PR}^{QS}$ can be easily derived from the equations of the corresponding intermediates in the original xTC equations, as shown in \Cref{tab:xtc_deriv}

\begin{figure*}
    \includegraphics[width=\textwidth]{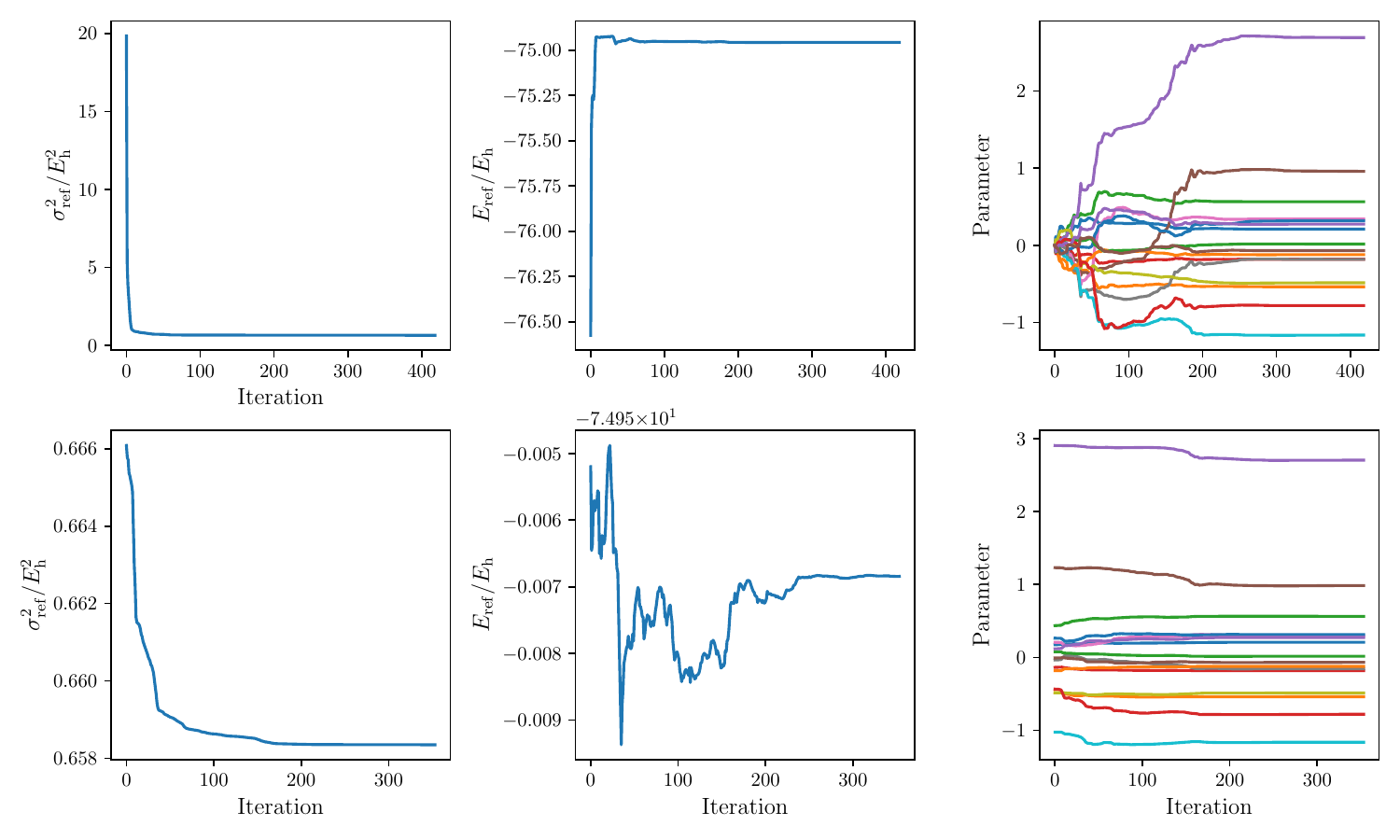}
    \caption{Variation of the variance of the reference energy, the reference energy and the various parameters in the Jastrow over the course of deterministic optimisation for the oxygen atom in the cc-pCVTZ basis set. This particular Jastrow was of the DTN form, with 16 parameters in total (4 e--e terms, 4 e--n terms and 8 e--e--n terms).  The top panels correspond to optimisation from all-zero parameters, while the bottom panels correspond to optimisation from a VMC guess of the parameters. It is clear that, despite the two very different starting points, the Jastrow parameters converge to very similar values.}
    \label{fig:optimisation}
\end{figure*}

\begin{figure*}
    \centering
    \includegraphics[width=\linewidth]{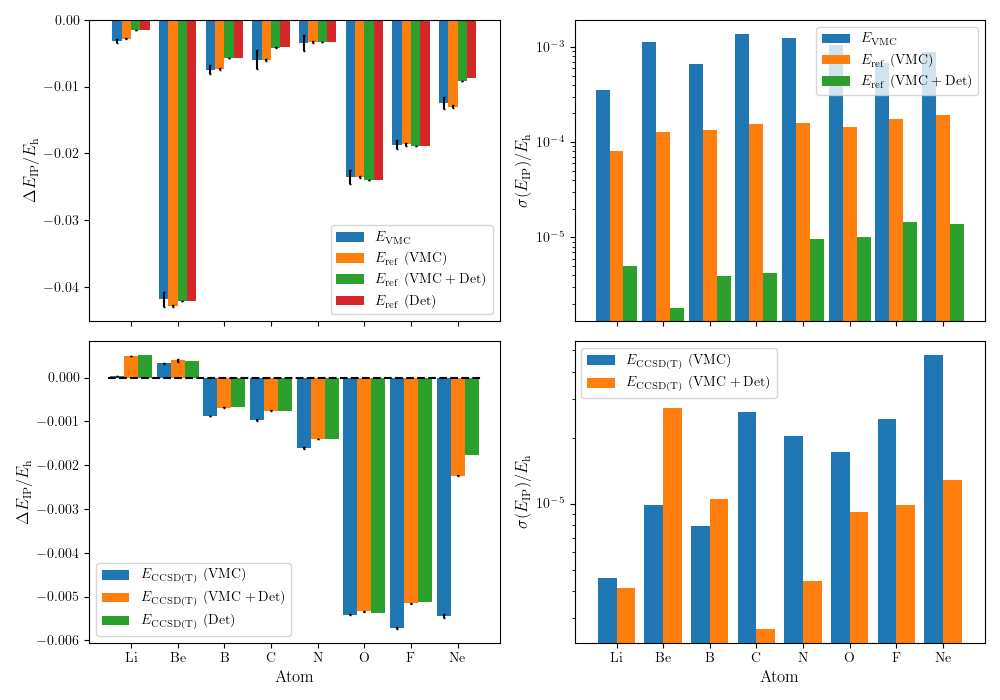}
    \caption{The error in the IPs of 1st row elements as computed using VMC and reference energies (top left), the corresponding standard deviations (top right), the error in the IPs of 1st row elements as computed using xTC-CCSD(T) (bottom left) and the corresponding standard deviations (bottom right). Standard deviations are obtained from five independent VMC runs for each species. All results are obtained from the cc-pCVTZ basis set. The information in brackets denotes the optimisation scheme employed.}
    \label{fig:1st_row_ips}
\end{figure*}

The full $\frac{d \Delta U_{PR}^{QS}}{df_l}$ term can then be obtained as
\begin{align}
    \begin{split}
      \frac{d \Delta U_{PR}^{QS}}{df_l} = &-\mathcal{P}^{(RS)}_{(PQ)} \Big( \rho_P^Q(i)\frac{d A_R^S(i)}{df_l} \\
      &+ \frac{d \mathbf{V}_P^Q(i)}{df_l} \cdot \mathbf{B}_R^S(i) + \mathbf{V}_P^Q(i) \cdot\frac{d \mathbf{B}_R^S(i)}{df_l}\Big).
    \end{split}
\end{align}

This is then used to compute the corresponding one- and zero-body corrections,

\begin{equation}
    \frac{d\Delta h_P^Q}{df_l} = - \frac{1}{2}\Big(\frac{d \Delta U_{PR}^{QS}}{df_l} - \frac{d \Delta U_{PR}^{SQ}}{df_l}\Big)\gamma_S^R
\end{equation}
and
\begin{equation}
    \frac{d\braket{\Phi_0|\hat L|\Phi_0}}{df_l} = -\frac{1}{3}\frac{d\Delta h_P^Q}{df_l} \gamma_Q^P
\end{equation}

\section{Results}
All VMC calculations reported in this section have been carried out using the CASINO package.\cite{Needs2020} Initial Hartree--Fock wavefunctions and integrals were obtained from PySCF,\cite{Sun2018} while corresponding TC integral computation and deterministic Jastrow optimisation were carried out using the TCHInt library and its Python interface, PyTCHInt.\cite{tchint} Integrals were computed using numerical quadrature over the direct product of atom-centered grids matching those used in PySCF. For the Jastrow optimisation, grid level 1 was used to reduce integral cost, while grid level 2 was employed for the final energy estimations. To perform a CCSD(T) calculation with the xTC Hamiltonian, we used the ElemCo.jl package\cite{elemcoil} to carry out a pseudocanonical $\mathrm{\lambda}$CCSD(T) calculation using biorthogonal orbitals,\cite{Kats2024} which we will refer to as xTC-CCSD(T) for the remainder of this work. 
\subsection{Basis set requirements}
\begin{figure*}
    \centering
    \includegraphics[width=\linewidth]{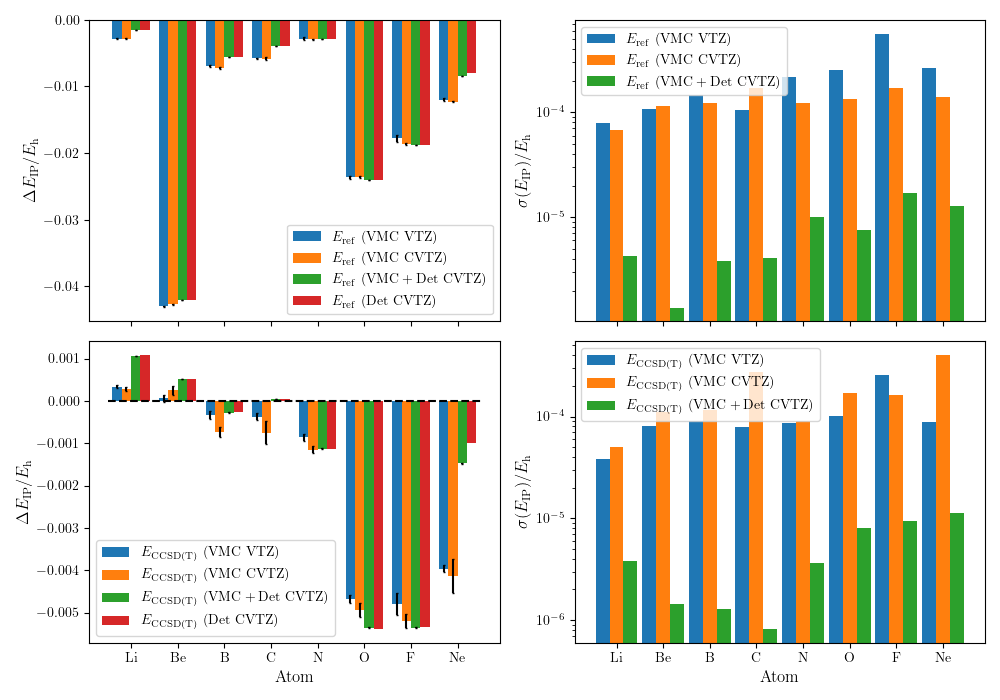}
    \caption{The error in the IPs of 1st row elements as computed using reference energies (top left), the corresponding standard deviations (top right), the error in the IPs of 1st row elements as computed using TC-CCSD(T) (bottom left) and the corresponding standard deviations (bottom right). Standard deviations are obtained from five independent VMC runs for each species. Blue bars correspond to results obtained from Jastrow factors optimised by VMC in the cc-pVTZ basis, while all others use the cc-pCVTZ basis for VMC and deterministic optimisation. Reference and TC-CCSD(T) energies were computed in the cc-pVTZ basis. The information in brackets denotes the optimisation scheme employed and the basis set used for final energy calculations.}
    \label{fig:1st_row_ips_tz}
\end{figure*}

The first important aspect to consider for this optimisation scheme is the dependence of its performance on the basis set used. VMC optimised Jastrow factors are expected to be relatively independent of the basis set, in particular when the Hartree--Fock determinant is used as the reference wavefunction, as this is usually well described even in small basis sets. However, by introducing an incomplete basis in \Cref{eq:var_resolution_0}, this invariance is lost in the deterministic optimisation scheme.

To demonstrate the effect of this we consider the Li atom in basis sets from the cc-pVXZ and cc-pCVXZ (X = D, T, Q) families\cite{Prascher2011a} and two simple BH Jastrow factors with a single optimisable parameter, corresponding to either an e--e or e-n term. The reference energies and their variances as a function of the Jastrow parameter, as computed using \Cref{eq:var_resolution_0} or \Cref{eq:variance_reference_vmc} are given in \Cref{fig:one_param}. It is clear that, while cc-pVXZ and cc-pCVXZ calculations lead to very similar VMC variances, this is not the case for the deterministic approach. For the cc-pVXZ basis sets, while the variance increases with basis set cardinality, it is significantly underestimated relative to VMC, while predicting a significantly shifted parameter value for the minimum. This effect is present in both the electron-electron and electron-nucleus Jastrow cases, albeit more pronounced in the former and can be ascribed to basis set incompleteness errors. The cc-pCVXZ basis sets show significant improvement, with cc-pCVTZ results agreeing with the VMC results in both magnitude and the position of the minimum. The inclusion of additional orbitals with significant core character is therefore crucial for a correct estimate of the variance of the reference energy and all deterministic Jastrow optimisations performed in the rest of this paper use basis sets of cc-pCVTZ quality. We note that, while the variance ranges over three orders of magnitude with different basis sets, the reference energies agree very well between basis sets. This suggests that while a relatively large basis set must be used to optimise the Jastrow factor, results are relatively transferable, and much smaller basis sets could be used for post-Hartree--Fock correlation treatments. We explore this hypothesis further for more complex systems and Jastrow factors in \Cref{sec:fr,sec:frm}.

\subsection{Deterministic Jastrow Optimisation}
For all results given below, the deterministic optimisation of the linear parameters in the Jastrow factor is done by minimising the variance computed according to \Cref{eq:var_resolution_0}, using the xTC approximation for the transcorrelated Hamiltonian integrals. The optimisation is carried out using the L-BFGS algorithm.\cite{Broyden70,Fletcher70,Goldfarb70,Shanno70,Nocedal1989} \Cref{fig:optimisation} shows an example of the convergence of the optimisation algorithm with L-BFGS iteration for the O atom, comparing a pure deterministic optimisation calculation to one initialised with VMC values for the Jastrow parameters. As expected, the latter starts much closer to the final result and overall they converge to very similar solutions for the Jastrow factor. 

\subsection{First row ionisation potentials}
\label{sec:fr}
\begin{table*}[!]
    \centering
    \begin{tabular}{c|c|c|c|c|c|c|c}
       MOLECULE  &  Total energy (HEAT) &  VMC (VTZ) & VMC(CVTZ) & VMC (aVTZ) & VMC + det (VTZ) & VMC + det (CVTZ) & VMC + det (aVTZ) \\
       \hline
        H&-0.500022&0.000124&-&0.000109&0.000126&-&$-0.00002$\\
        C&-37.845311&0.004715&0.003252&0.003851&0.002811&0.002639 &0.001947\\
        N&-54.589743&0.008351&0.006626&0.006508&0.006510&0.005674&0.004698\\
        O&-75.068210&0.016152&0.012040&0.015061&0.014952&0.013496&0.010903\\
        F&-99.735148&0.023856&0.018668&0.023393&0.023495&0.020955&0.018150\\
        \hline
        N$_2$&-109.543481&0.022666&0.021423&0.017381&0.021194 &0.01939&0.015987\\
        O$_2$&-150.328884&0.036044&0.033140&0.026535&0.030875 &0.026545&0.022136\\
        F$_2$&-199.532431&0.047067&0.045415&0.032684&0.040114 &0.037278&0.027948 \\
        \hline
        HF&-100.460908&0.027603&0.027335&0.018526&0.029173&0.026783&0.01963 \\
        H$_2$O&-76.439478&0.021737&0.021283&0.013207&0.022366 &0.021093&0.013878\\
        \hline
        CO&-113.327394&0.023659&0.021483&0.018297&0.020992 &0.018820&0.016119 \\
        \hline
    \end{tabular}
    \caption{Errors in the total energies of first row atoms and diatomics, computed using xTC-CCSD(T) with Jastrow factors optimised by VMC or by combined VMC and deterministic optimisation, in Hartree. Total HEAT values for the energies are given in the second column. Calculations were carried out using cc-pVTZ (VTZ), cc-pCVTZ (CVTZ) and aug-cc-pVTZ (aVTZ) basis sets. For the VMC-only calculations, Jastrow optimisation was carried out in the same basis set as the post--HF calculation. For VMC followed by deterministic optimisation (VMC + det), both phases of Jastrow optimisation were carried out in the cc-pCVTZ basis, regardless of the basis used for the post-HF calculation.}
    \label{tab:diatomics}
\end{table*}
We consider the first ionisation potentials of atoms
from the first row of the periodic table, Li--Ne. For each atom and ion, we consider two optimisation routines, applied to a DTN Jastrow factor with $N_u = N_\chi = 4$ and $N_f = 2$: (a) we directly optimise the Jastrow factor deterministically; and (b) we initially do a VMC optimisation, followed by a refinement of the Jastrow factor by deterministic optimisation. We set the number of VMC configurations to target a standard deviation in $E_\mathrm{ref}$ of 0.1 mHartree. The deterministic optimisation is carried out until the variance of the reference energy is converged to within $10^{-6}$ Hartree$^2$, and we expect the observed standard deviations could be further reduced by employing a more stringent convergence criterion. In \Cref{fig:1st_row_ips} we present ionisation potential errors and standard deviations obtained at the reference energy level and using xTC-CCSD(T).\cite{Kats2024} We find that reference energies obtained by either the (a) or (b) optimisation routines generally agree with those from the purely stochastic optimisation, although for some systems (Li, Al, Ne) they exhibit lower errors. Significantly, the standard deviation of the reference energies decreases by more than an order of magnitude when deterministic optimisation is used to refine the VMC-optimised Jastrow factors

At the xTC-CCSD(T) level, the decrease in variance is less pronounced but increases with system size. However, we once again observe a lowering of the error in the xTC-CCSD(T) IPs, which is as high as a factor of three for the Ne ionisation.

The previous results are obtained using the cc-pCVTZ basis set\cite{Dunning1989a, Woon1995a} for both Jastrow optimisation and the post-Hartree--Fock correlation treatment. We assess the transferability of the deterministically optimised Jastrow factors by using them to obtain reference and xTC-CCSD(T) energies in the cc-pVTZ basis set,\cite{Dunning1989a} as shown in \Cref{fig:1st_row_ips_tz}. In general, the results are comparable with those obtained from Jastrow factors optimised directly by VMC in the cc-pVTZ basis, with notable decreases in variance when employing deterministically optimised Jastrow factors. This suggests that post-Hartree-Fock calculations can be safely carried out in smaller basis sets than those used for Jastrow optimisation, avoiding prohibitive scaling with basis set size.

\subsection{First-row molecules}
\label{sec:frm}

We consider first-row homonuclear diatomics N$_2$, O$_2$ and F$_2$, and heteronuclear diatomic CO, and assess the performance of deterministically optimised Jastrows for the dissociation energy of these systems, relative to pure VMC optimisation. For more complex Jastrow factors, the latter has previously been shown to be sufficient to accurately describe these molecular properties using aug-cc-pVTZ basis sets.\cite{Dunning1989a,Kendall1992a} As a proof of concept, we use here the same Jastrow parametrisation from the previous section, which has fewer terms than the one used in Ref. \onlinecite{Christlmaier2023}, which results in less accurate results than those reported there. Total energies obtained by xTC-CCSD(T) are shown in \Cref{tab:diatomics}, with the corresponding atomisation energies in \Cref{tab:diatomics-atom}. 

\begin{table*}[]
    \centering
    \begin{tabular}{c|c|c|c|c|c|c|c}
       MOLECULE  &  Atom. energy (HEAT) &  VMC (VTZ) & VMC(CVTZ)  & VMC + det (VTZ) & VMC + det (CVTZ) & VMC (aVTZ) & VMC + det (aVTZ)\\
       \hline
        N$_2$&0.363995&-0.005964&-0.008171&-0.008174&-0.008042&-0.004365&-0.006591\\
        O$_2$&0.192464&-0.003700&-0.003018&-0.000971&0.000447&-0.002455&-0.000330 \\
        F$_2$&0.062135&0.000765&0.001371&0.006876&0.004632&0.004652&0.008352\\
        \hline
        HF&0.225638&-0.003623&-0.003818&-0.005552&-0.005702&0.000251&-0.0015\\
        H$_2$O&0.871246&-0.005337&-0.005874&-0.007162&-0.007345&-0.000949&-0.003015\\
        \hline
        CO&0.413873&-0.002792&-0.003170&-0.003229&-0.002685&-0.002406&-0.003269\\
    \hline    
    \end{tabular}
    \caption{Errors in the atomisation energies of first row molecules, computed using xTC-CCSD(T) with Jastrow factors optimised by VMC or by combined VMC and deterministic optimisation, in Hartree. HEAT values for the atomisation energies are given in the second column. Calculations were carried out using cc-pVTZ (VTZ), cc-pCVTZ (CVTZ) and aug-cc-pVTZ (aVTZ) basis sets. For the VMC-only calculations, Jastrow optimisation was carried out in the same basis set as the post--HF calculation. For VMC followed by deterministic optimisation (VMC + det), both phases of Jastrow optimisation were carried out in the cc-pCVTZ basis, regardless of the basis used for the post-HF calculation.}
    \label{tab:diatomics-atom}
\end{table*}

We carry out calculations in the cc-pVTZ, cc-pCVTZ and aug-cc-pVTZ basis sets. Deterministic optimisation of the Jastrow factor is always carried out in the cc-pCVTZ basis set, irrespective of which basis set is then used for the xTC-CCSD(T) calculation. For pure VMC optimisation of the Jastrow factor, the target xTC-CCSD(T) basis set is employed. In all cases deterministic optimisation reduces the error in the total energies significantly, although the cancellation of errors sometimes proves less advantageous for the atomisation energy. 

We also consider two hydrides, HF and H$_2$O. The deterministic optimisation routine employed in this paper generally requires core-like functions in the basis set used for Jastrow optimisation. However, no such basis sets are available for the hydrogen atom, as it only has valence electrons. We find that in this case using the cc-pVTZ basis set for hydrogen in the optimisation does not introduce significant error relative to the energy obtained by VMC optimisation. Therefore, for these molecules we optimise the Jastrow factor using a cc-pCVTZ basis on the heavy atom and cc-pVTZ on hydrogen, with results also given in \Cref{tab:diatomics} and \Cref{tab:diatomics-atom}. We find that, in this case, deterministic optimisation only reduces the total energy error in the core-valence basis set, and error cancellation in the atomisation energy is generally poorer for the deterministically optimised Jastrow factors. Nevertheless, in most cases results are of a similar quality to those obtained by pure VMC optimisation of the Jastrow factors and we expect that sufficiently flexible Jastrow parametrisations would be able to recover the sub-milliHartree accuracy of previously reported results, in a more reliable, noise-free way.

\subsection{Deterministic optimisation for variational energies}

\begin{figure}
    \centering
    \includegraphics[trim=0.5cm 0 0 0,clip, width=0.5\textwidth]{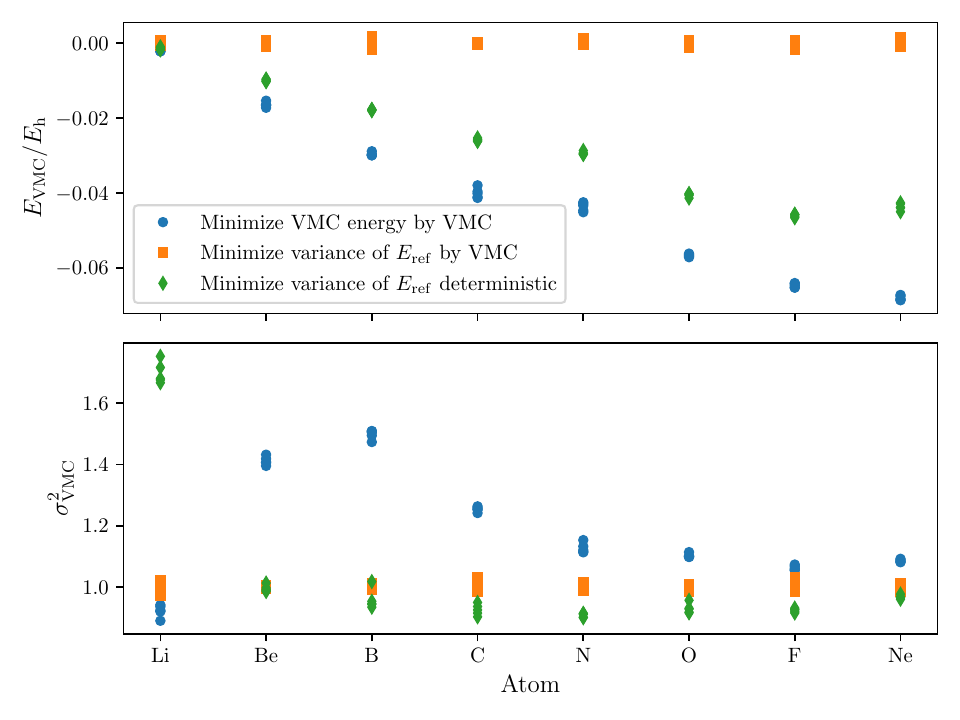}
    \caption{VMC energies and variances obtained by variational energy minimisation (blue), variance of the reference energy minimisation (orange) and the proposed deterministic minimisation protocol (green). Energies were shifted such that the mean of the $E_\mathrm{ref}$ variance minimisation corresponds to 0. The variances were scaled such that the mean of the $E_\mathrm{ref}$ variance minimisation corresponds to 1.  All VMC calculations used the cc-pVTZ basis set for the Hartree--Fock distribution.}
    \label{fig:vmc}
\end{figure}

Finally, we consider the quality of the variational energies of Slater--Jastrow wavefunctions obtained using deterministic optimisation of the Jastrow factor, without further applying the transcorrelated methodology. \Cref{fig:vmc} shows the VMC energies $E_\mathrm{VMC}$ and their corresponding variances $\sigma_\mathrm{VMC}^2$ for first row atoms, computed by VMC minimisation of either $E_\mathrm{VMC}$ or $\sigma_\mathrm{ref}^2$, as well as those obtained by enhancing VMC minimisation of $\sigma_\mathrm{ref}^2$ with deterministic optimisation. Corresponding average values of $E_\mathrm{VMC}$ and $\sigma_\mathrm{VMC}^2$ are given in \Cref{tab:vmc}.

We find that the variational energies obtained by employing deterministic minimisation are significantly lower than those from VMC optimisation of $\sigma_\mathrm{ref}^2$, approaching those obtained by direct variational energy minimisation. However, $\sigma_\mathrm{VMC}^2$ remains much lower than the values obtained from energy minimisation, suggesting this approach as a viable scheme to obtain VMC energies that are both variationally close to the ground state and have low variances, and thereby have utility beyond TC applications.

\begin{table}[h]
    \centering
    \begin{tabular}{c|c|c|c|c|c|c|}
     \multirow{2}{*}{Atom} &\multicolumn{2}{c|}{Min. $E_\mathrm{VMC}$}&\multicolumn{2}{c|}{Min. $\sigma_\mathrm{ref}^2$} &\multicolumn{2}{c|}{Min. determ. $\sigma_\mathrm{ref}^2$}\\
     \cline{2-7}
  &$E_\mathrm{VMC}$&$\sigma_\mathrm{VMC}^2$&$E_\mathrm{VMC}$&$\sigma_\mathrm{VMC}^2$&$E_\mathrm{VMC}$&$\sigma_\mathrm{VMC}^2$\\
  \hline
  Li & -7.4765&0.0132&-7.4748&0.0143&-7.4763&0.0243\\
  Be & -14.6381 & 0.1074 & -14.6216  &0.0761 & -14.6316 &0.0758 \\
  B & -24.6172&0.4027 & -24.5875 & 0.2689&-24.6054&0.2580\\
  C & -37.8040 & 0.7121& -37.7639 & 0.5679& -37.7897 &0.5261 \\
  N & -54.5461 & 1.3954 &-54.5024 &1.2382 & -54.5317 &1.1248\\
  O & -75.0123 & 2.7475 & -74.9563 & 2.4914& -74.9969 &2.3180\\
  F & -99.6715 & 5.0309& -99.6068 &4.7323 & -99.6529 &4.3727 \\
  Ne & -128.8676 &8.7779 & -128.7995 & 8.0831& -128.8432& 7.8219\\
  \hline
    \end{tabular}
    \caption{Mean VMC energies and variances obtained by variational energy minimisation, variance of the reference energy minimisation and the proposed deterministic minimisation protocol. Energies are given in Hartree and variances in Hartree$^2$. All VMC calculations used the cc-pVTZ basis set for the Hartree--Fock distribution.}
    \label{tab:vmc}
\end{table}
\section{Conclusions}

In this paper, we have presented a new deterministic Jastrow optimisation scheme based around the minimisation of the variance of the transcorrelated reference energy, as computed in a finite basis set. Core-like basis functions are crucial for the success of this approach, as in their absence the computed variance is significantly underestimated, and the wrong parametrisation is predicted for the minimum.

We benchmark this approach against the ionisation potentials of first-row atoms, as well as the total energies and atomisation energies of a representative set of second-row molecules. In general, results are of a similar standard to those obtained by VMC optimisation. We find that, when used in conjunction with VMC, deterministic optimisation significantly reduces the variance of the estimated reference energies, as well as the resulting xTC-CCSD(T) energies. For total molecular energies, it also results in lowered errors, although error cancellation in the atomisation energies is slightly poorer in most cases, with the notable exception of the oxygen molecule. 

The reduction of variance achievable through this deterministic optimisation scheme opens up new directions for transcorrelated approaches, most notably the treatment of weak interactions, such as dispersion. In this case, the noise due to the VMC optimisation would be sufficient to occlude any interaction, so we are interested to explore the performance of deterministically optimised Jastrow factors in this setting in a future publication.

Recently, the use of effective core potentials (ECPs) in conjunction with transcorrelation has been proposed as a means to reduce the overhead of such calculations by reducing the required basis set size, while maintaining and in some cases improving accuracy.\cite{Simula2024} This reduction would be beneficial for deterministic optimisation as well, however, we need to explore how this approach can be effectively combined with deterministic Jastrow optimisation, particularly due to its effective removal of core orbitals from the wavefunction. In a similar direction, it may be possible to carry out the all-electron calculation, but perform the Jastrow optimisation using a frozen-core xTC approach to compute the variance. In the present work, we find that orbitals with significant density in the core region are required for the optimisation, but the important contributions come from virtual orbitals, not so much from the occupied core orbitals, so ECP and frozen core approaches may be viable pathways to reducing the cost of deterministic Jastrow optimisation.

Finally, this approach is promising for pure VMC applications as well, as it provides a better trade-off between VMC energy and variance than conventional VMC approaches based on direct energy or variance minimisation, which tend to generate large values of whichever quantity is not being minimised. VMC optimisation of the variance of the TC reference energy already provided an improvement in this case, but this is further enhanced by the use of the deterministic optimisation scheme proposed in this work.

\section{Acknowledegments}
The authors gratefully acknowledge funding from the Max Planck Society. 

\nocite{*}
\bibliography{bibliography}
\end{document}